\documentclass{article}
\usepackage{spconf,amsmath,graphicx}
\usepackage{url}
\usepackage{hyperref}
\usepackage{booktabs}       
\usepackage{tabularx} 
\usepackage{threeparttable}
\usepackage{stfloats}
\usepackage{multirow}
\usepackage{makecell}
\usepackage{enumitem}
\usepackage{color}
\usepackage[bottom]{footmisc}

\title{UNIT-DSR: DYSARTHRIC SPEECH RECONSTRUCTION SYSTEM USING SPEECH UNIT NORMALIZATION}

\name{Yuejiao Wang$^1$, Xixin Wu$^{1,2}$, Disong Wang$^2$, Lingwei Meng$^1$, Helen Meng$^{1,2}$ \vspace{-1.0em}}

\address{
$^1$ Department of Systems Engineering and Engineering Management,\\
The Chinese University of Hong Kong, Hong Kong SAR, China \\
  $^2$ Vocal Engineering Technologies Limited, Hong Kong SAR, China\\
  \texttt{\ninept{\{wangy, wuxx, dswang, lmeng, hmmeng\}@se.cuhk.edu.hk}} 
  }

\begin{document}
\ninept
\maketitle
\begin{abstract}
Dysarthric speech reconstruction (DSR) systems aim to automatically convert dysarthric speech into normal-sounding speech. The technology eases communication with speakers affected by the neuromotor disorder and enhances their social inclusion. NED-based (Neural Encoder-Decoder) systems have significantly improved the intelligibility of the reconstructed speech as compared with GAN-based (Generative Adversarial Network) approaches, but the approach is still limited by training inefficiency caused by the cascaded pipeline and auxiliary tasks of the content encoder, which may in turn affect the quality of reconstruction. Inspired by self-supervised speech representation learning and discrete speech units, we propose a Unit-DSR system, which harnesses the powerful domain-adaptation capacity of HuBERT for training efficiency improvement and utilizes speech units to constrain the dysarthric content restoration in a discrete linguistic space. Compared with NED approaches, the Unit-DSR system only consists of a speech unit normalizer and a Unit HiFi-GAN vocoder, which is considerably simpler without cascaded sub-modules or auxiliary tasks. Results on the UASpeech corpus indicate that Unit-DSR outperforms competitive baselines in terms of content restoration, reaching a 28.2\% relative average word error rate reduction when compared to original dysarthric speech, and shows robustness against speed perturbation and noise\footnote[1]{\href{https://wyj1996.github.io/Unit-DSR-demo/index.html}{Demo page: https://wyj1996.github.io/Unit-DSR-demo/index.html} }.

\end{abstract}

\section{INTRODUCTION}
\label{sec:intro}
Dysarthric speech with abnormal acoustic characteristics \cite{knuijt2014dysarthria} can lead to the deterioration of the speech-motor control system, which engenders communication barriers for dysarthria patients. This study introduces an innovative approach to the dysarthric speech reconstruction (DSR) task, aiming to transform dysarthric speech into a more natural and intelligible form.

Existing DSR systems are primarily based on voice conversion (VC), which can be divided into two popular frameworks: generative adversarial network (GAN) and neural encoder-decoder (NED). GAN regards DSR as a cross-domain shift problem by directly mapping dysarthric speech features to their normal counterparts, and different variants have been investigated, e.g., CycleGAN \cite{imai2020improving}, MaskCycleGAN \cite{prananta22_interspeech} and DiscoGAN \cite{purohit2020intelligibility}; NED designs separate encoders for content restoration, prosody correction, and speaker identity preservation, respectively, followed by a decoder and vocoder for normal speech generation \cite{wang2020end, wang2022speaker}. The drawback of GAN-based DSR systems is that the linguistic and prosody aspects of speech are modeled implicitly, which makes the reconstruction errors difficult to trace. However, NED explicitly reconstructs different speech components through encoders, offering higher interpretability and controllability.

Cascaded NED-based DSR systems typically employ a content encoder to extract linguistic representations required for healthy speech generation. To ensure precise extraction of linguistic content from dysarthric speech, various auxiliary tasks have been investigated: E2E-DSR \cite{wang2020end} force-aligns its content encoder with the text encoder of a text-to-speech (TTS) model using cross-modal knowledge distillation; ASA-DSR \cite{wang2022speaker} fine-tunes a well-trained automatic speech recognition (ASR) model as the content encoder to extract phonetic posteriorgrams as linguistic representations; a series of Parrotron systems \cite{biadsy2019parrotron, chen21w_interspeech, doshi2021extending} convert dysarthric spectrograms to their normal version, with phoneme recognition added as an auxiliary task to constrict the content encoder output within the linguistic content space. Given the limited dysarthric speech data and intra-patient variance in dysarthria symptoms, these systems are often trained with multiple objectives on large-scale normal speech to constrain content encoder outputs within the linguistic space and then adapted to dysarthric subjects with limited utterances.

However, the cascaded pipeline of NED-based systems encounter several issues: 1) Training inefficiency: the modules within the pipeline necessitate training from scratch across multiple phases, followed by fine-tuning towards dysarthric subjects, which is complicated and inefficient; 2) Sub-optimal performance: the content restoration heavily relies on the content encoder that is simultaneously optimized with auxiliary tasks, e.g., ASR \cite{wang2022speaker, chen21w_interspeech} and modal alignment \cite{wang2020end}. Although guiding the encoder to better capture the linguistic content, the auxiliary tasks distract the optimization from the ultimate goal of speech reconstruction and lead to sub-optimal final performance. Consequently, imprecise linguistic information from the content encoder can be propagated through the decoder and vocoder, further exacerbating content restoration.


To alleviate these issues, we draw inspiration from self- supervised learning (SSL) of speech representations and discrete speech units, which have achieved success in voice conversion (VC) \cite{polyak2021speech, tsai2022superb}, ASR, and textless speech translation \cite{lee2021textless, nguyen2023improving}. 1) SSL representations reveal powerful few-shot domain adaptation capabilities. SSL speech models, such as HuBERT \cite{hsu2021hubert} and Wav2Vec 2.0 \cite{baevski2020wav2vec}, have demonstrated effectiveness in discriminative and generative tasks to tackle domain shift problems \cite{yang2021superb, tsai2022superb}. Domain-adapted Wav2Vec 2.0 representations \cite{hu2023exploring} have also been successfully integrated into automatic dysarthric speech recognition; 2) Discrete speech units, clustered from SSL speech representations, are proved to be highly related with phonemes while weakly correlated with speaker or gender \cite{sicherman2023analysing}, forming a discrete space with less para-linguistic information. Therefore, speech units often serve as linguistic representations in VC \cite{polyak2021speech} and offer a shortcut for bypassing text and mel-spectrogram aspects in speech translation \cite{nguyen2023improving}.

In this study, we propose a Unit-DSR system based on the HuBERT model and discrete speech units. The system comprises two modules -- a speech unit normalizer and a Unit HiFi-GAN vocoder -- striving to convert diverse dysarthric pronunciation patterns into their normal versions of a reference speaker, which is the essence of `normalization'. The normalizer, initialized from the HuBERT model, first transforms dysarthric speech into a healthy speech unit sequence using the connectionist temporal classification (CTC) loss and a multi-stage fine-tuning strategy. Subsequently, the Unit HiFi-GAN vocoder generates waveforms directly from speech units.
The contributions of the work include: 1) We are among the first to introduce speech units as discrete content representations in the DSR task, which avoids auxiliary tasks and outperforms content encoder outputs employed in previous works \cite{wang2020end, wang2022speaker}; 2) We propose to utilize the HuBERT backbone with an effective multi-stage fine-tuning strategy, which showcases its powerful domain adaptation capability and significantly enhances training efficiency; 3) The overall Unit-DSR structure is greatly simplified without cascaded sub-modules or complex training objectives, while still outperforms competitive baselines on the UASpeech corpus.

\vspace{-0.4em}
\section{PROPOSED METHOD}
\label{sec:methods}

\begin{figure*}[htb]

  \centering
  \includegraphics[width=0.86\textwidth]{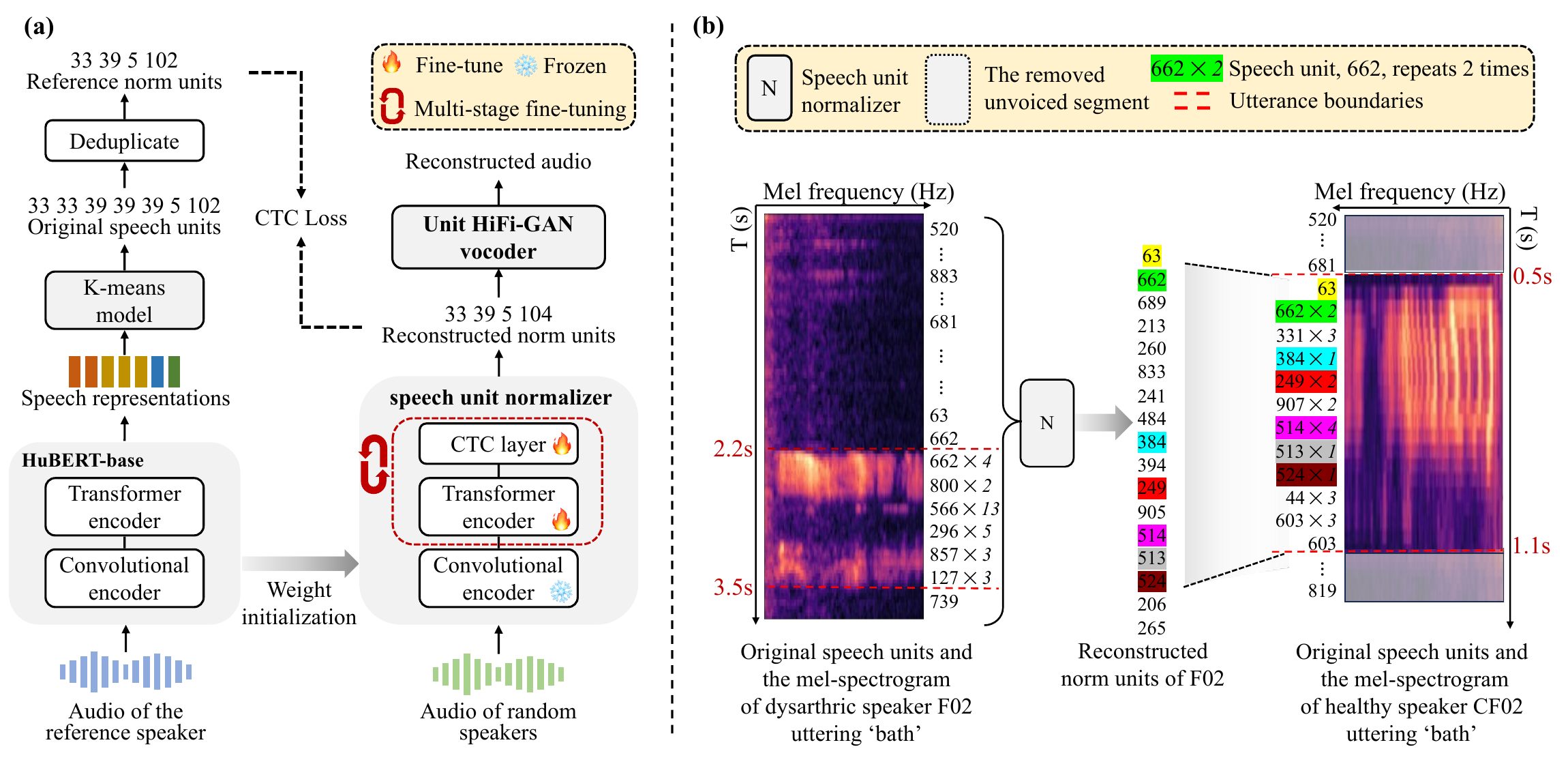}
  \vspace{-0.3em}
  \caption{(a) Diagram of the Unit-DSR system. (b)  An example of original speech units of different speakers uttering `bath', and the reconstructed norm units from the speech unit normalizer, which have a high correspondence with the reference speech units.}
  \label{unit-dsr structure}
  
  \vspace{-0.9em}
\end{figure*}

The Unit-DSR system, as shown in Fig. \ref{unit-dsr structure}(a), is composed of a speech unit normalizer and a Unit HiFi-GAN vocoder \cite{lee2021textless}. First, a pre-trained HuBERT produces the reference speech units and initializes the speech unit normalizer, followed by a multi-stage fine-tuning. The Unit HiFi-GAN vocoder is modified from HiFi-GAN vocoder and enhanced with a unit-duration predictor.

\vspace{-0.7em}
\subsection{HuBERT and speech units}


HuBERT is a readily available SSL speech representation model that iteratively optimizes a BERT-like loss based on the K-means clusters of the model's intermediate representations. HuBERT comprises a convolutional encoder, a transformer encoder, a projection layer, and a code embedding layer. By applying the prediction loss exclusively to the masked regions, HuBERT learns a unified acoustic and language model from continuous inputs \cite{hsu2021hubert}.


After pre-training, the learned K-means clusters can be utilized to discretize the input audio's intermediate representations into a sequence of cluster indices, represented as $[z_1, z_2, ..., z_T], z_i \in [0, 1, 2, ..., K-1]$, where $T$ and $K$ denote the number of frames and clusters, respectively. These cluster indices serve as the original speech units. The speech units are generated at a 20ms frame rate for audio sampled at 16kHz and contain many repeated units within the sequence. Analysis reveals a strong correlation between speech units and phoneme families, while their association with speaker or gender is weak \cite{sicherman2023analysing}. Consequently, speech units can effectively constrain the speech content component in a discrete space, and their duration can be explicitly controlled by modifying the repetitions.

\vspace{-0.7em}
\subsection{Speech unit normalizer}

Speech units from audios containing the same content can exhibit significant variations due to factors such as speaking styles, emotions, silence, and background noise. Additionally, the variance is aggravated by the severity levels and pathologies of dysarthric patients. As demonstrated in the example of Fig. \ref{unit-dsr structure}(b), the speech units within the utterance boundaries of F02 and CF02 are quite different.

Therefore, to transform inaccurate contents and modify inappropriate pauses of dysarthric speech into a normal pattern in the form of speech units, a speech unit normalizer is built. In Fig. \ref{unit-dsr structure}(a), the normalizer is initialized using a pre-trained HuBERT model with a CTC layer and then fine-tuned in a multi-stage way, using the speech of various speakers (healthy or dysarthric) as input and norm units of a healthy reference speaker as the target.

The left branch in Fig. \ref{unit-dsr structure}(a) depicts the process of the reference norm units extraction, while the right branch is the fine-tuning and inference pipeline of the Unit-DSR system. First, pairs of utterances from a healthy reference speaker and a random speaker (either healthy or dysarthric, depending on the fine-tuning stage) in the same content are required. Unvoiced segments on both sides of the reference utterances are removed. Next, in the left branch, speech representations of the 11th-layer of HuBERT are extracted from the reference speech, and the K-means model clusters these representations into original speech units. To ease the learning burden on the normalizer and eliminate the duration information, duplicate units in the original speech units are removed. Therefore, we get the reference norm units, which serve as the target for the fine-tuning stages. 

Simultaneously, in the right branch, the utterance of a random speaker with the same content is fed into the normalizer. After passing through the HuBERT and a CTC layer, the waveform is transformed into a reconstructed norm-unit sequence. Specifically, the transformer encoder and the CTC layer are fine-tuned using the CTC loss in a multi-stage strategy (details in subsection \ref{multi-stage}). This process converts the reconstructed norm units into the reference norm units, thereby normalizing various acoustic patterns, including dysarthric patterns, into the healthy pattern of the reference speaker. Fig. \ref{unit-dsr structure}(b) provides an example of the reconstructed norm units and their strong correspondence with the reference speech units. Finally, the reconstructed norm units are put into the Unit HiFi-GAN vocoder to generate the reconstructed waveform directly (details in subsection \ref{vocoder section}).

\vspace{-0.7em}
\subsection{Multi-stage fine-tuning strategy}
\label{multi-stage}

To effectively adapt the initial speech unit normalizer into a DSR-oriented normalizer, we propose a multi-stage fine-tuning strategy. It begins by fine-tuning the normalizer using a small amount of healthy speech to obtain a typical speech normalizer, which is then adapted using the dysarthria dataset. Different stages are distinguished by the variations in the reference-random speaker pairs:

\begin{itemize}[leftmargin=0.3cm, itemindent=0.2cm]
\vspace{-0.2em}
\item \textbf{First stage}: we use multiple speakers from the VoxPopuli ASR dataset \cite{wang2021voxpopuli} as random speakers and the speaker of LJSpeech \cite{ljspeech17} as the reference speaker. Since the content of VoxPopuli and LJSpeech is not parallel, a text-to-unit conversion is performed using a transformer machine translate model trained on LJSpeech with characters as input and norm-unit as target (refer to \cite{lee2021textless} for more details). It enables us to get utterance-reference norm-unit pairs. At this stage, the normalizer learns the basic normalization rules, but can not transfer the ability well to different reference speakers and corpus, e.g., the noisy UASpeech corpus \cite{kim2008dysarthric}.

\vspace{-0.1em}
\item \textbf{Second stage}: a healthy speaker from the dysarthria dataset is selected as the new reference speaker. The random speakers are the other healthy speakers within the dysarthria dataset, and all of their utterances are used for fine-tuning. At this stage, the speech unit normalizer can better adapt to the patterns of the new reference speaker and the dysarthria corpus.
\vspace{-0.1em}
\item \textbf{Third stage}: a dysarthric patient is chosen as the random speaker, and the reference speaker remains the same as in the second stage. Only utterances of this patient are used for fine-tuning, allowing the normalizer to remove the dysarthric pattern.
\vspace{-0.2em}
\end{itemize}

\vspace{-0.9em}
\subsection{Unit HiFi-GAN vocoder}
\label{vocoder section}

\begin{figure}[tb]
  \centering
  \includegraphics[width=0.87\linewidth]{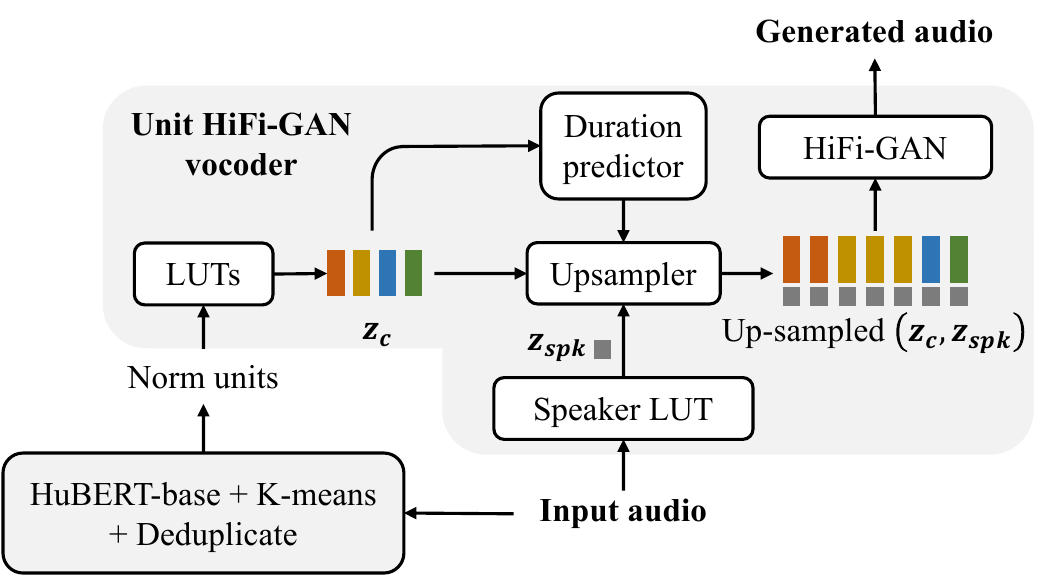}
  
  \caption{The structure of Unit HiFi-GAN vocoder.}
  \label{vocoder}
  \vspace{-1.5em}
\end{figure}

Our multi-speaker Unit HiFi-GAN vocoder directly decodes waveform from the reconstructed norm units, as shown in Fig. \ref{vocoder}. It is optimized by the generator-discriminator loss and the mean square error of the predicted duration of each unit in the logarithmic domain. First, the norm-unit sequence is converted into representations $z_c$ via look-up tables (LUTs). The representation is then up-sampled according to the unit duration obtained from the duration predictor. The speaker embedding, $z_{spk}$, is extracted from a speaker LUT and concatenated to each frame of the up-sampled $z_c$. Finally, the generator accepts the up-sampled representations of $(z_c, z_{spk})$ and converts them into a waveform.

\begin{table}[tbp]
  \scriptsize
  \caption{The dataset distributions of the multi-stage fine-tuning strategy in our experimental setup. UA. is the UASpeech corpus.}
  \vspace{0.3em}
  \label{duration distributions}
  \centering
  
\begin{threeparttable}  
  \begin{tabular}{c|ccc}
    \toprule
    \makecell[c]{Stages} &   \makecell[c]{Dataset of the\\reference speaker} &  \makecell[c]{Dataset of\\random speakers} &\makecell[c]{Total durations\\of random speakers}\\
    \midrule
    1 & LJSpeech & VoxPopuli ASR & 10.0 h\\
    2 &UA. (CF02) & UA. (healthy speakers) & 11.9 h\\
    3 &UA. (CF02) & UA. (M05/F04/M07/F02) & 1.2 h\\
    
    \bottomrule
  \end{tabular}
\end{threeparttable} 
\vspace{-1.6em}
\end{table}

\section{EXPERIMENTAL SETUP}

The Unit-DSR system is evaluated using the UASpeech corpus \cite{kim2008dysarthric}, one of the largest databases for English dysarthria, which consists of 16 dysarthric speakers and 13 normal speakers. The speakers were asked to repeat 455 unique words, including 155 common words and 300 uncommon words. These words were divided into three blocks: B1-B3. In this study, four patients with varying severity levels, M05 (middle), F04 (middle), M07 (low), and F02 (low), are selected for subjective and objective evaluations.

During the model pre-training, the speech normalizer is initialized by an open-sourced multi-lingual HuBERT-base model, which is trained on VoxPopuli unlabeled speech with 4.5k hrs of data for English, Spanish, and French, respectively. Speech units are extracted via K-means clustering (K=1000) with features from the 11th layer of the third iteration (follow the setting in \cite{lee2021textless}). 
For the multi-stage fine-tuning, the dataset distributions are outlined in Table \ref{duration distributions}. The first stage setting has been explained in subsection \ref{multi-stage}, and an open-sourced LJSpeech-normalizer for the first stage can be found in \cite{lee2021textless}.  In the second (or third) stages, only B1 and B3 of healthy speakers (or the dysarthric speaker) are employed for fine-tuning, with B2 reserved for testing. We perform 10k updates and use Adam optimization in all stages. And Detailed HuBERT fine-tuning parameters can be found in fairseq.
The Unit HiFi-GAN is trained using the LJspeech and UASpeech corpora, taking two days on 2 GTX A6000 GPUs. The code for the Unit HiFi-GAN  is released \cite{polyak2021speech}.

Three baseline systems are applied for comparison: one baseline is the ASR-TTS system, which consists of a multi-lingual HuBERT-CTC ASR model and an open-sourced TTS model (Tacotron 2 \cite{shen2018natural}), with a well-trained HiFi-GAN vocoder. The architecture and the fine-tuning dataset of its ASR model are identical to those of the Unit-DSR system; the other two baselines include the E2E-DSR system \cite{wang2020end} and the ASA-DSR system \cite{wang2022speaker}, which both follow the NED pipeline and have a larger training set than Table \ref{duration distributions}. 

\vspace{-0.5em}
\section{RESULTS}

\subsection{Baselines comparison on content restoration}
\label{Baseline comparison}

The 5-scale mean opinion score (MOS) test and ASR are conducted to evaluate the content restoration accuracy. In the subjective MOS test, 20 listeners are invited to assess 15 random-selected words from the B2 set, focusing on the content similarity between the reconstructed speech and the reference speech of CF02, while disregarding speaker identity and background noise. In the objective evaluation, the publicly available ASR system, Jasper \cite{li2019jasper} with greedy decoding, is applied to get the word error rate (WER) of the B2 set.

\begin{table}[tbp]
  \scriptsize
  \caption{The 5-scale MOS test scores for content similarity with mean scores and the 95\% confidence intervals.}
  \vspace{0.4em}
  \label{mos_test}
  \centering
  
\begin{threeparttable}  
  \begin{tabular}{c|cccc}
    \toprule
    Systems & M05 & F04 & M07 & F02  \\
    \midrule
    Original  & 2.86 $\pm$ 0.15  & 2.42 $\pm$ 0.14 & 1.75 $\pm$ 0.11 & 1.95 $\pm$ 0.12  \\
    ASR-TTS  & 3.45 $\pm$ 0.12  & 3.21 $\pm$ 0.13 & 3.35 $\pm$ 0.13 & 3.02 $\pm$ 0.15  \\
    E2E-DSR  & 3.61 $\pm$ 0.14  & 3.47 $\pm$ 0.14 & 3.72 $\pm$ 0.18 & 3.87 $\pm$ 0.16  \\
    ASA-DSR  & 4.19 $\pm$ 0.10  & 3.98 $\pm$ 0.11 & 3.71 $\pm$ 0.17 & 4.26 $\pm$ 0.14  \\
    Unit-DSR  & \textbf{4.52 $\pm$ 0.14}  & \textbf{4.65 $\pm$ 0.11} & \textbf{4.62 $\pm$ 0.11} & \textbf{4.55 $\pm$ 0.12}  \\
    
    \bottomrule
  \end{tabular}
\end{threeparttable} 
\vspace{-1em}
\end{table}

\begin{table}[tbp]
\scriptsize
  \caption{WER (\%) comparison for UASpeech. $\Delta$ refers to the percentage decrease in WER compared to original dysarthric speech.}
  \vspace{0.3em}
  \label{ASR_result}
  \centering
\begin{threeparttable}  
  \begin{tabular}{c|cccc}
    \toprule
    Systems & \makecell[c]{WER of\\M05 / $\Delta$} & \makecell[c]{WER of\\F04 / $\Delta$} & \makecell[c]{WER of\\M07 / $\Delta$} & \makecell[c]{WER of\\F02 / $\Delta$}  \\
    \midrule
    Original  & 81.7/ --  & 81.7/ -- & 95.6/ -- & 95.9/ --  \\
    ASR-TTS  & 74.2/ -18.5\%  & 75.4/ -7.7\% & 70.0/ -26.8\% & 81.6/ -14.9\%  \\
    E2E-DSR  & 69.8/ -23.3\%  & 69.3/ -15.2\% & 73.1/ -23.5\% & 72.0/ -24.9\%  \\
    ASA-DSR  & \textbf{62.5/ -31.1\%} & 65.6/ -19.7\%  & 62.7/ -34.4\% & \textbf{65.8/ -31.4\%}  \\
    Unit-DSR  & 64.4/ -29.2\%  & \textbf{65.5/ -19.8\%} & \textbf{62.1/ -35.0\%} & 68.3/ -28.8\%  \\
    
    \bottomrule
  \end{tabular}
\end{threeparttable} 
\vspace{-1em}
\end{table}

As shown in Table \ref{mos_test}, the proposed Unit-DSR achieves higher content similarity, indicating the effectiveness of speech units in reconstructing speech with accurate content and pronunciations. In contrast, the content encoders of other baselines expose more phoneme errors. Readers are encouraged to refer to the demo page for further details. The ASR results are summarized in Table \ref{ASR_result}, where Unit-DSR achieves the highest relative WER reductions of 19.8\% and 35.0\% on F04 and M07, respectively. It also demonstrates comparable WER results with ASA-DSR on M05 and F02. However, even built upon the same backbone of the Unit-DSR, the ASR-TTS system still performs the worst. We assume that content errors at the character level are more easily amplified by the decoder and vocoder than errors at the unit level (also indicated in Fig. \ref{unit-dsr structure}(b)). These results imply that our Unit-DSR system can attain higher content restoration accuracy using a simplified pipeline, without an auxiliary task for the content encoder. 

Since Jasper is not as robust as humans and can be easily influenced by noise or speaking styles, we further conduct a human listening test to evaluate the intelligibility of the reconstructed speech by Unit-DSR. Two listeners are invited to judge whether the utterance corresponds to the word label. And this human listening test is conducted in the subsequent ablation study and robustness study, based on 50 common words in the B2 set.

\vspace{-0.7em}
\subsection{Ablation study on multi-stage fine-tuning}

To evaluate the effectiveness of the multi-stage fine-tuning strategy and assess the contribution of different stages, we design an ablation study based on the metric of human listening test accuracy. Four versions of the Unit-DSR system, each fine-tuned with various combinations of stages, are compared in Table \ref{ablation}.

\begin{table}[htbp]
\scriptsize
  \caption{Human listening test accuracy (\%) for the ablation study on multi-stage fine-tuning strategy. $\Delta$ refers to the percentage of accuracy decrease compared with the original Unit-DSR system.}
  \vspace{0.3em}
  \label{ablation}
  \centering

\resizebox{\linewidth}{!}{
\begin{threeparttable}  
  \begin{tabular}{l|cccc|c}
    \toprule
    \makecell[c]{Stages} & M05 / $\Delta$ & F04 / $\Delta$ & M07 / $\Delta$ & F02 / $\Delta$ & Ave. $\Delta$ \\
    \midrule
    1 + 2 + 3  & 82.0/ --  & 72.0/ -- & 82.0/ -- & 76.0/ -- & -- \\
    1 + 3  & 79.0/ -3.7\%  & 66.0/ -8.3\% & 66.0/ -19.5\% & 70.0/ -7.9\% & -9.9\% \\
    1 + 2 & 14.0/ -82.9\%  & 48.0/ -33.3\% & 6.0/ -92.7\% & 16.0/ -78.9\% & -72.0\% \\
    1 & 8.0/ -90.2\% & 10.0/ -86.1\% & 2.0/ -97.6\% & 4.0/ -94.7\% & -92.2\% \\
    
    \bottomrule
  \end{tabular}
   \vspace{-1.2em}
\end{threeparttable} 
  }

\end{table}

We find that the fine-tuning strategy is essential for adapting the pre-trained HuBERT model to the DSR task. Not surprisingly, the dysarthric speaker in stage 3 contributes the most (72.0\%) to the average accuracy. However, healthy speakers from UASpeech also significantly contribute to the perception accuracy: adding stage 2 to stage 1 lead to a 20.2\% increase in average accuracy, and the Unit-DSR without stage 2 suffers an average accuracy loss of 9.9\%. These effects are particularly obvious for speakers F04 and M07.

\vspace{-0.7em}
\subsection{Robustness study}

\begin{figure}[htbp]
  \centering
  \includegraphics[width=\linewidth]{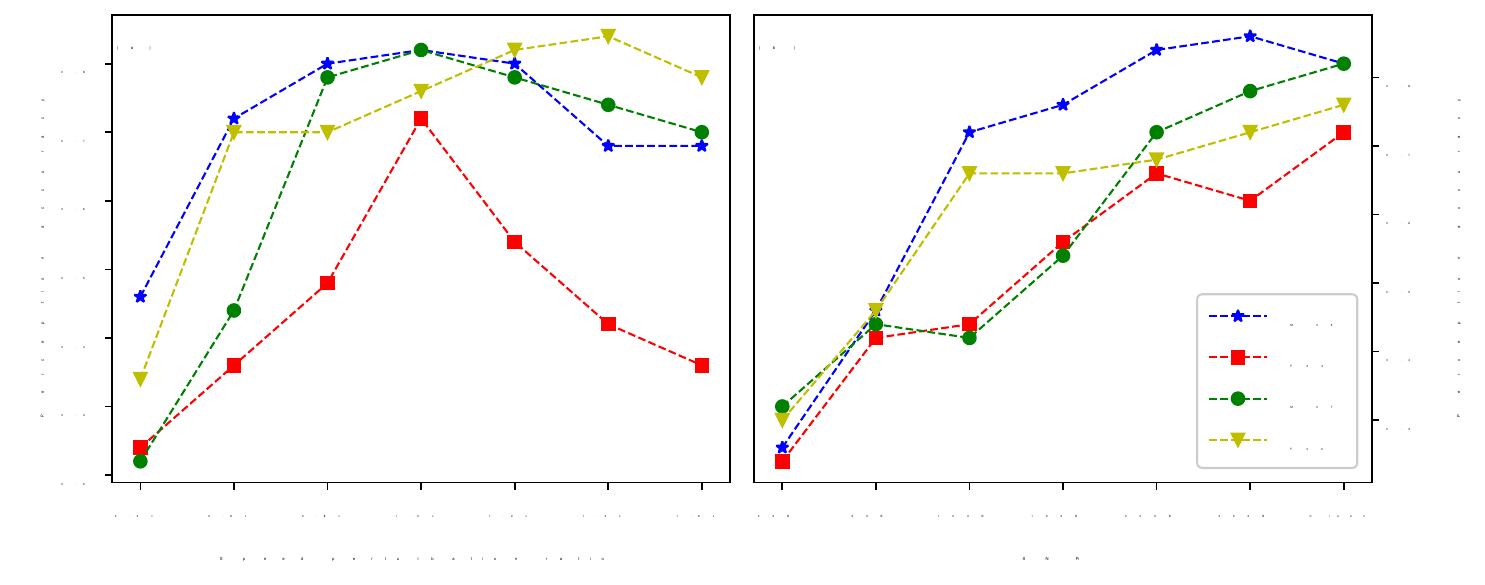}
  \vspace{-2.0em}
  
  \caption{Human listening test accuracy of Unit-DSR with (a) different input speed perturbation ratios and (b) various input SNRs.}
  \label{robust}
\vspace{-1em}
\end{figure}

The robustness study examines how the reconstructed norm units are influenced by the distribution fluctuations of input audios. We conduct speed perturbation on dysarthric speech to simulate changes in the patient's speaking rate and add Gaussian white noise at different power levels to represent various record conditions. Their human listening test results are depicted in Fig. \ref{robust}.

The Unit-DSR system encodes accurate content with speed perturbation ratios ranging from 0.8 to 1.6, except for speaker F04. When speech is slowed down by a rate of 0.4 or 0.6, many unnatural utterances will be interpolated, causing the normalizer to generate incorrect norm- units. Speaker F04 is a unique case because her utterances are quite rushed, often omitting stresses and pauses, and a higher speed perturbation ratio will lost some crucial pronunciation cues.
Similarly, in Fig. \ref{robust}(b), when the SNR falls below 10 dB, important phoneme features cannot be reconstructed due to noise. However, when the SNR is above 15 dB, the performance of the model remains relatively stable. In conclusion, Unit-DSR demonstrates robustness against distribution fluctuations in input audios.
\vspace{-0.8em}

\vspace{-0.6em}
\section{CONCLUSIONS}
\vspace{-0.2em}
The proposed unit-DSR system normalizes the dysarthric speech into a reference speaker and generates speech directly through speech units. It omits the cascaded pipeline while improve the intelligibility significantly. And the pretrained Hubert model and multi-stage fine-tuning strategy helps to efficiently adapt to various severity-level patients. Combined with multi-speaker unit-vocoder, unit-DSR can also generate speech with different timbres. Furthermore, our model has the potential to extend  to dysarthria corpus in other textless languages based on speech units.

\vspace{-0.5em}
\section{ACKNOWLEDGEMENTS}

This research is supported by the Centre for Perceptual and Interactive Intelligence and the CUHK Stanley Ho Big Data Decision Analytics Research Centre.


\vfill

\clearpage

\bibliographystyle{IEEEbib}
\ninept
\bibliography{strings,refs}

\end{document}